# Large area graphene synthesis on catalytic copper foil by an indigenous electron cyclotron resonance plasma enhanced chemical vapor deposition setup

S. Karmakar,[1] S. K. Kundu,[1] and G. S. Taki[1,a]

[1]*Department of Electronics and Communication Engineering, Institute of Engineering & Management, Kolkata, 700091, India*

A novel route for the synthesis of large area graphene by an indigenously built electron cyclotron resonance (ECR) plasma enhanced chemical vapor deposition (PE-CVD) setup has been reported in this work. A unique 2.45 GHz permanent magnet type ECR PE-CVD system has been designed and developed for the growth of various nano-structures and films. The apparatus has a unique advantage to synthesize large area graphene at much lower additional substrate heating than heat supplied by an external oven in conventional thermal CVD method. A major amount of heat is supplied here by ECR plasma. Polycrystalline copper foil is used here as a catalytic substrate which is pre annealed inside resonant hydrogen plasma prior to its exposure under precursor's gaseous plasma inside the ECR PE-CVD reactor. Methane is used as carbon precursor along with hydrogen as carrier gas. Argon gas is used for the rapid cooling of the substrate maintaining suitable thermodynamic condition favorable for graphene synthesis. The synthesis process has been carried out by controlling growth temperature, total pressure and synthesis time at various gas ratios. The exposed copper foil has been characterized by Raman spectroscopy to investigate the quality of synthesized graphene in the home made reactor.

## I. INTRODUCTION

Graphene is an atomically thick layer of carbon atoms arranged in a hexagonal honeycomb lattice structure. In general, graphene is a two-dimensional (2D) allotrope of carbon, the fourth most abundant element in the universe by mass, and considered as a wonder material owing to its astonishing physical, electrical, mechanical, optical, thermal and chemical properties that concentrates enormous interest towards numerous technological and fundamental studies. The graphene was theoretically studied by physicist P. R. Wallace[1] in 1947 and its 2D structure has been finally synthesized, utilizing mechanical exfoliation of graphite, by A. Geim and K. Novoselov[2] in 2004 that led them to receive the Nobel Prize in physics 2010. The synthesis of graphene is one of the most challenging research topics over the past decade. The researchers are developing graphene films mostly by physical and chemical methods. The chemical methods are most commonly used to prepare graphene oxide (GO) and reduced graphene oxide (rGO). Among all, the thermal chemical vapor deposition (CVD) technique[3-5] is much preferred owing to its cost effectiveness, large area synthesis and superior quality. But a large complicated heating oven is essential for thermal CVD reactor. A special Plasma Enhanced CVD (PE-CVD) technique has been employed here for rapid synthesis of graphene at much less heat supply to the substrate. This is possible due to the fact that ECR resonance plasma heats up the substrate to a large extent.

___

[a]gstaki@iemcal.com

Conventionally, the thermal chemical vapor deposition (CVD) is preferred for large area production of monolayer and multi-layer graphene films on transition metal (e.g., copper, nickel, cobalt etc.) catalytic substrate[6-11]. It has been observed that the nucleation and growth kinetics of graphene on crystalline transition metal surface requires very high growth temperature ~600°C-1000°C[12]. Usually, in thermal CVD, the hydrocarbon gas precursors (e.g., methane, acetylene etc.) gets decomposed into carbon atoms and other fragments at higher temperature before dissolving into the hot catalytic surface leading to the formation of graphene during the cooling phase[13-16]. The mechanism for synthesizing graphene in an in-house developed 2.45 GHz ECR PE-CVD reactor[17-19] and the investigation of its optical properties will be discussed here.

## II. EXPERIMENTAL METHOD

### A. Significance of ECR PE-CVD based graphene synthesis

In this work, an Electron Cyclotron Resonance (ECR) plasma surface of the in-house developed unique 2.45 GHz ECR PE-CVD reactor plays a major role for graphene synthesis. The ECR plasma is extremely suitable for supplying enormous thermal energy to the precursor and also providing sufficient heating to catalytic substrate. Unlike the thermal CVD, methane ($CH_4$) gas molecules get dissociated to form radicals and ions obtaining high energy stored at the resonance plane of ECR plasma. The most difficult external convectional heating used in thermal CVD is not needed in this case. In this advantageous catalyst heating method, the polycrystalline copper foil is placed at the vicinity of ECR resonance zone. In addition, an additional substrate heating arrangement has been provided inside the reactor chamber for achieving the desired thermodynamic condition considering various thickness and size of the catalytic substrate. An enormous stored energy of ECR plasma raises the foil at a temperature of 300°C, measured by an in-house developed copper-constantan thermocouple. The process minimizes the excess substrate heating temperature as low as ~300°C to attain uniform growth temperature for graphene synthesis. A DC regulated substrate heating assembly, made up of Nichrome filament, has been used to provide complementary heating up to a maximum of 600°C inside the chamber, keeping under vacuum. Less costly polycrystalline copper foils were used as catalytic substrate considering its low solubility of carbon compared to any other transition metal and also for the easy availability of its chemical etchant. In this synthesis method, several parameters have been investigated e.g., growth time, substrate temperature, precursor/carrier gas ratio and total pressure of the reactor chamber. Each of these parameters has important roles in graphene formation. A schematic view of the experimental set up has been presented in figure 1.

### B. Thermodynamic considerations for growth process

Precise control of various growth parameters are essential for synthesis of pristine graphene and the deposition of graphene from methane ($CH_4$) gas may be expressed by an equation given below:

$$CH_4(g) \rightarrow C_{graphite} + 2H_2(g) \tag{1}$$

Considering the crystalline copper foil as an ideal catalytic substrate, the partial pressure calculation of methane ($CH_4$) inside the ECR PE-CVD system will provide the insight on super-saturation of carbon for monolayer graphene growth. The free energy of graphene reaction for this research may be expressed by

$$\Delta G = \Delta G^0 + RT \ln(P_{CH_4}/P_{H_2}^2) \tag{2}$$

Here, $\Delta G$ is the change in free energy and can be defined as $G_{reactants} - G_{products}$. This free energy change is considered in convention with crystal growth. Super-saturation of monolayer graphene will be realized if $\Delta G$ value is positive but at equilibrium, $\Delta G$ value is zero. $\Delta G^0$ is defined as the change in free energy at standard state. $P_{CH_4}$ and $P_{H_2}$ are the respective partial pressure of methane and hydrogen in the reactor chamber during synthesis. R is the universal gas constant whereas, T indicates the growth temperature. The reaction entropy and enthalpy of the decomposition reaction of methane towards graphite at a particular temperature can be calculated from the thermodynamic datasheet[20]. This is important to understand the decomposition reaction towards graphene production and also to calculate the value of $\Delta G^0$. Calculating the value of $\Delta G^0$ at growth temperature of T=700°C, the partial pressure ratio between methane and hydrogen needs to be maintained at $P_{CH_4}: P_{H_2} = 1: 130$ for monolayer graphene synthesis.

**C. Synthesis process**

The surface of 20 mm×20 mm size 50 μm thick commercially available copper foil is preprocessed by filtered and diluted solution of lime extract (contains ~8% citric acid). This process makes the copper surface clean from contamination and washed in running de-ionized (DI) water for a considerable time. The substrate pieces are dried and preserved in an inert atmosphere. During the synthesis process, multiple copper foils are placed on top of a stainless steel substrate mounting table attached to the substrate manipulating shaft. An alumina disk provides the insulation between the substrate mounting table and the manipulating shaft. The substrate can be accurately positioned at the vicinity of ECR resonance zone with the help of the manipulator. The temperature of the substrate is directly measured by a copper-constantan thermocouple meter. A 100 liters/min anti-suck-back double stage rotary vane pump creates the vacuum inside the reactor chamber. An ultimate vacuum pressure level of 5×10⁻³ mbar is achieved and measured with the help of a Pirani vacuum gauge controller. 2.45 GHz, 700 watt microwave power has been injected into the CVD chamber near 0.0875 Tesla iso-gauss surface to create stable ECR plasma. This plasma provides sufficient energy to dissociate precursor gas molecules. Initially, hydrogen ($H_2$) is injected to raise the pressure up to 5.5×10⁻² mbar to ignite the plasma. The hydrogen plasma is used to etch the catalytic polycrystalline copper substrate in presence of additional internal heating up to 700°C. The process helps to mitigate surface irregularities and improves grain boundaries suitable for graphene growth. The process also creates a reducing environment inside the chamber to prevent the foil from oxidation. Hence, methane ($CH_4$) gas is injected to maintain a gas ratio of $H_2:CH_4$ at 10:1 after etching. The growth times are varied at discrete values of 5, 10, 15 and 20 minutes respectively. During this period, the dissociated carbon species (e.g., $\dot{C}H_3$, $\dot{C}H_2$, $\dot{C}H$ etc.) are allowed to deposit onto the catalytic substrate and further catalytic decomposition takes place as represented in figure 2. Microwave power is switched off to cease the growth process. The complementary heating and the flow of methane & hydrogen have also been stopped simultaneously. Argon (Ar) is allowed to enter into the reactor chamber up to 5 mbar for cooling the catalytic substrate at a cooling rate of ~ 4°C/min. During this cooling period, the solubility of carbon into copper substrate is decreased and carbon atoms float over the copper surface forming hexagonal bonds of graphene. The synthesis mechanism illustrating process time and temperature are represented graphically in figure 3. Exposed copper is taken out from deposition chamber to carry out copper etching with the help of 0.25 M Ferric Chloride ($FeCl_3$) solution. After 3-4 Hours, copper is fully dissolved into the etchant solution and the graphene layer floats on top of the etchant solution as shown in figure 4. Repeated water replacement technique utilizing DI water minimizes the $FeCl_3$ contamination on the graphene layer. The graphene film is finally transferred onto a silicon wafer.

## D. RESULTS AND DISCUSSIONS

The exposed copper has been characterized by Raman Spectroscopy, under laser excitation of 633 nm, to ensure the quality of graphene. Figure 5 shows Raman Spectra of the graphene film grown for 5 minutes on the commercially available copper foil at a growth temperature of 700°C and controlled vacuum pressure of $6 \times 10^{-2}$ mbar. The overall appropriate growth temperature is achieved by the combination of plasma heating and complementary additional heating. It causes the formation of hexagonal cluster of graphene on the catalytic copper substrate at the suitable thermodynamic condition. Raman spectra show three peaks of D band (at ~1380 $cm^{-1}$), G band (at ~1620 $cm^{-1}$) and 2D band (at ~2680 $cm^{-1}$). The peaks are marginally shifted due to the rise in temperature of the sample, caused by incident laser beam. D band is considered as disordered band indicating structural defects, edge defects and dangling $sp^2$ carbon bonds that breaks the symmetry in graphene structure. The G band indicates the degree of graphitization. The value of D/G intensity ratio calculated from Raman Spectra is 0.73. The lower D/G value indicates about less oxidation between $sp^2$ hybridized carbon bonds that leads to pristine graphene. The characteristic peak of graphene structure is denoted by 2D band. The intensity ratio 2D/G determines the number of graphene layers. The intensity ratio ~2–3 denotes monolayer graphene, 2>2D/G>1 denotes bilayer graphene and 2D/G<1 signifies multilayer graphene structure. In this case, the value of 2D/G intensity ratio calculated from Raman Spectra is 0.42 indicating the development of multilayer graphene.

The optical microscopic image of previous sample presented in figure 6, clearly shows the formation of hexagonal clusters of graphene on exposed copper foil. The hexagonal shaped clusters are about 1 μm in dimension and overlapped each other as observed in optical microscope image. The formation of these clusters can also be seen for a similar sample characterized by Scanning Electron Microscope (SEM). Figure 7 represents the SEM image in which the darker patches are the indication that graphene clusters formed on the surface of exposed copper.

**III. CONCLUSIONS**

A novel internally heating method for synthesizing graphene has been proposed in this work. The ECR plasma provides enough heat energy to the copper substrate and additionally an auxiliary internal small heater provides low complementary heating power for the graphene growth. This ECR plasma assisted method finds its application for graphene synthesis over a large area catalytic substrate which has not been carried out and published so far. The characterization result of Raman Spectroscopy presented here, confirms about the graphene growth on copper substrate indicating D, G and 2D peak. The intensity ratio 2D/G<1 indicates the formation of multilayer graphene. The darker patches seen on the SEM image and the hexagonal cluster formation represented in optical microscopic image are the strong evidence of graphene developed on catalytic copper substrate. This preliminary study is essential for the optimization of growth parameters to obtain good quality large area pristine graphene. It can be observed from the partial pressure calculation of methane and hydrogen that at the operating growth temperature of 700˚C, a high gas ratio $P_{CH_4}:P_{H_2}$ =1:130 are to be maintained for development of monolayer graphene. Hence, a further development over the existing setup has been planned to minimize the contamination by lowering the background pressure utilizing ultra-high vacuum (UHV) pumping system. It will enhance the ECR plasma density and heating power to a greater extent to facilitate monolayer graphene synthesis.

**ACKNOWLEDGEMENT**

The authors would like to show their extreme gratitude to Dr. Satyajit Chakrabarti, Director, Institute of Engineering & Management (IEM), Kolkata for providing support to set up Applied Materials Research Laboratory (AMRL) at IEM, Kolkata. The authors are thankful to Dr. Malay Gangopadhyay, Head, Dept. of E.C.E., Institute of Engineering & Management, Kolkata for providing constant encouragement and support to carry out this research work. The authors would like to acknowledge The Institution of Engineers (India) for sanctioning a grant-in-aid (Project ID: RDPG2016055) to develop a 2.45 GHz ECR PE-CVD reactor system. The Raman Spectroscopy and optical microscopy are carried out at IISER, Kolkata for which the authors are grateful to Dr. Goutam Dev Mukherjee, IISER, Kolkata. Thanks are also due to Sayak Dutta Gupta for characterization by SEM at Micro Nano Characterization Facility, IISc Bangalore.

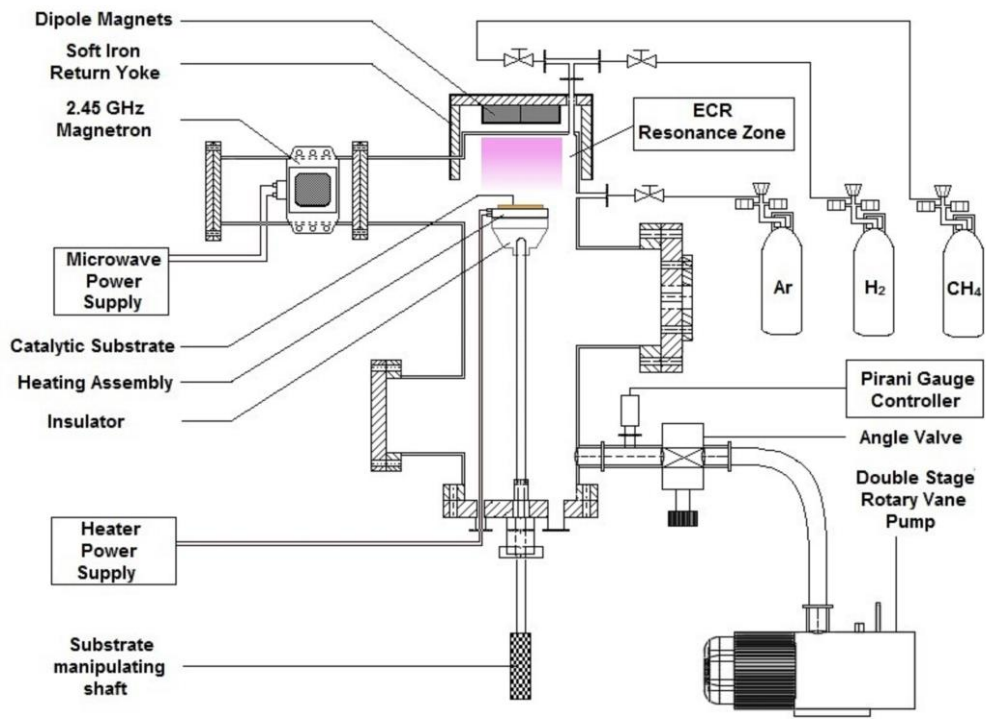

FIG. 1. Schematic of 2.45 GHz ECR PE-CVD system for graphene synthesis

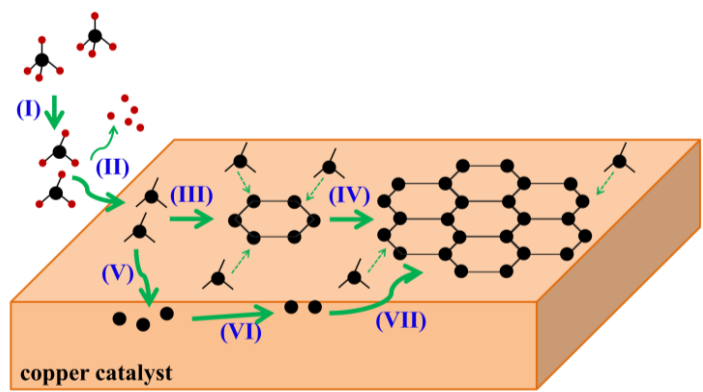

FIG. 2. Schematic diagram shows the surface kinetic processes of graphene formation on copper substrate from methane ($CH_4$).
        step (I)   : dehydrogenation ($CH_4 \rightarrow \dot{C}H_3$)
        step (II)  : further dehydrogenation and catalytic decomposition ($\dot{C}H_3 \rightarrow C$)
        step (III) : nucleation on catalytic copper surface
        step (IV) : graphene growth and cluster formation from surface mediated reaction
        step (V)  : diffusion of some carbon adatoms in copper lattice
        step (VI) : migration of diffused carbon to copper surface during cooling process
        step (VII): graphene growth and cluster formation from diffused carbon atoms

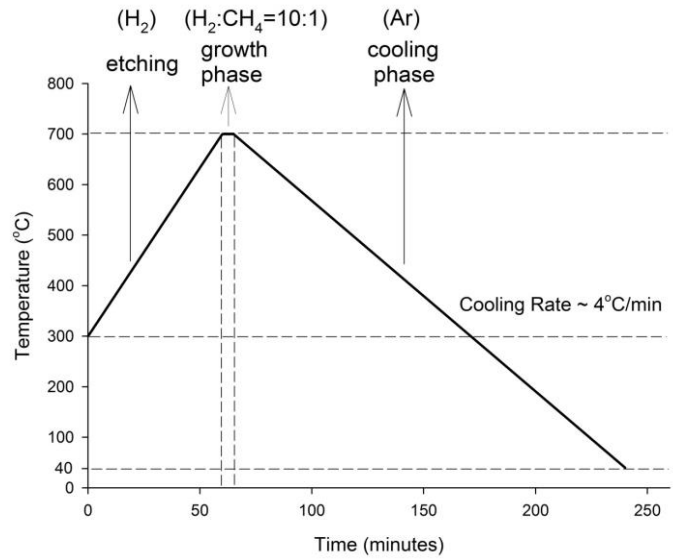

FIG. 3. Process flow of graphene synthesis

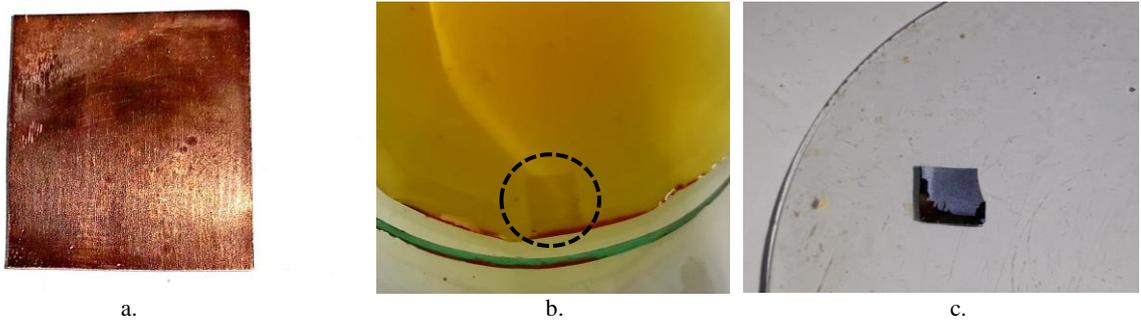

a.                            b.                            c.

FIG. 4. a. Optical image of the synthesized copper taken as catalytic substrate, b. Graphene film floating on $FeCl_3$ solution marked inside dotted ring, c. Graphene film transferred on silicon wafer

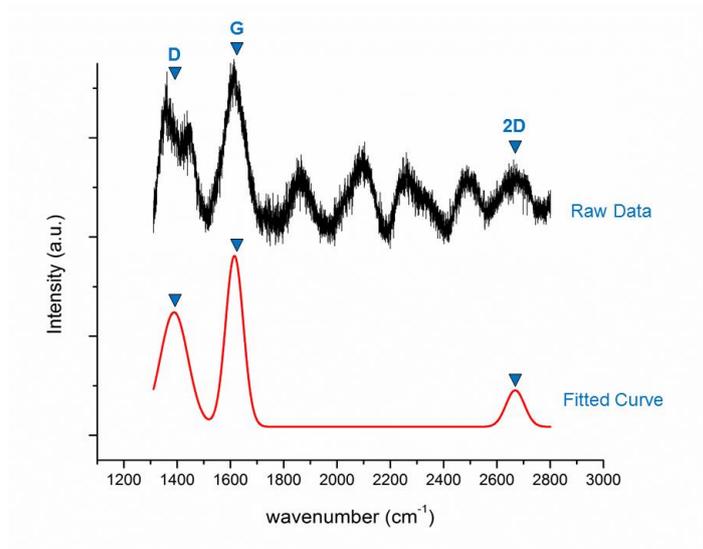

FIG. 5. Raman spectroscopy of graphene film grown on the surface of copper foil

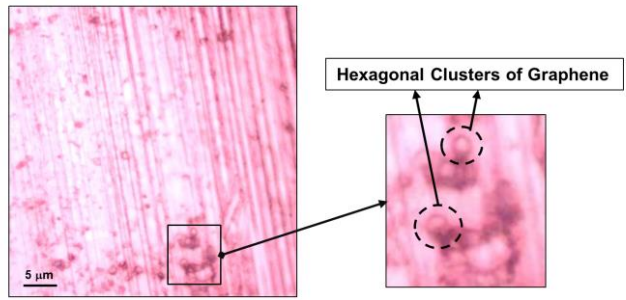

FIG. 6. Optical microscopic image of hexagonal cluster formation of graphene on surface of copper substrate

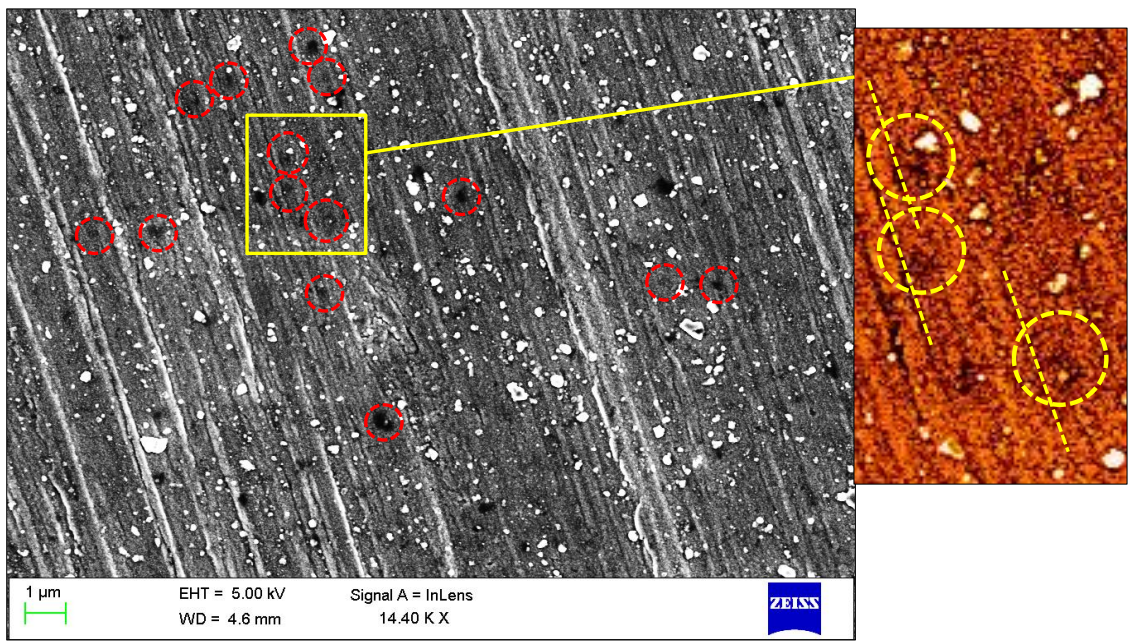

FIG. 7. SEM image indicating darker patches of deposited carbon at the surface of copper substrate

9